\documentclass[a4paper, 11pt]{article}

\usepackage[english]{babel}
\usepackage[utf8x]{inputenc}
\usepackage[T1]{fontenc}
\usepackage[authoryear]{natbib}
\usepackage{latexsym,amssymb,amsmath,graphicx,epsfig,amsthm}
\usepackage{bm}
\usepackage{algorithm}
\usepackage[noend]{algpseudocode}

\usepackage[a4paper,top=3cm,bottom=2cm,left=3cm,right=3cm,marginparwidth=1.75cm]{geometry}

\usepackage{amsmath}
\usepackage{graphicx}
\usepackage[colorinlistoftodos]{todonotes}
\usepackage[colorlinks=true, allcolors=blue]{hyperref}
\usepackage{subfig}

\title{Sensitivity of Codispersion to Noise and Error in Ecological and Environmental Data}
\author{Ronny Vallejos$^{*}$, Hannah L. Buckley$^{\dag}$, Bradley S. Case$^{\dag}$,\\ Jonathan Acosta$^{*}$, Aaron M. Ellison$^{\ddag}$}
\date{\small $^{*}$Departamento de Matem\'atica, Universidad T\'ecnica Federico Santa Mar\'ia, Valpara\'iso, Chile\\
$^{\dag}$ School of Science, Auckland University of Technology, Auckland, New Zealand \\
$^\ddag$Harvard Forest, Harvard University, Petersham, Massachusetts, USA
}

\begin{document}
\maketitle
%-------------------------------------------------------------------------------------------------
% ABSTRACT
%-------------------------------------------------------------------------------------------------

\begin{abstract}
Codispersion analysis is a new statistical method developed to assess spatial covariation between two spatial processes that may not be isotropic or stationary. Its application to anisotropic ecological datasets have provided new insights into mechanisms underlying observed patterns of species distributions and the relationship between individual species and underlying environmental gradients. However, the performance of the codispersion coefficient when there is noise or measurement error ("contamination") in the data has been addressed only theoretically. Here, we use Monte Carlo simulations and real datasets to investigate the sensitivity of codispersion to four types of contamination commonly seen in many real-world environmental and ecological studies. Three of these involved examining codispersion of a spatial dataset with a contaminated version of itself. The fourth examined differences in codisperson between plants and soil conditions, where the estimates of soil characteristics were based on complete or thinned datasets. In all cases, we found that estimates of codispersion were robust when contamination, such as data thinning, was relatively low (<15\%), but were sensitive to larger percentages of contamination. We also present a useful method for imputing missing spatial data and discuss several aspects of the codispersion coefficient when applied to noisy data to gain more insight about the performance of codispersion in practice. 
% * <hannah.buckley@aut.ac.nz> 2017-07-12T20:15:53.662Z:
% 
% > but they decrease 
% I think that this is unclear because we haven't said that to 'test' the effects of contamination on codispersion we took the approach of correlating a variable with a contaminated version of itself (for all but the BCI analysis). We either need another sentence or two here to more clearly state the approaches taken or we need to make this statement of results more generic. Perhaps we can say something like "We found that estimates of codispersion were robust when contamination was relatively low, but became affected when the percentage of contamination was greater than 20%." (probably need something more specific than 'became affected'...?)
% 
% ^ <aellison@fas.harvard.edu> 2017-07-30T21:27:08.077Z:
% 
% AME: inserted additional text and added values for relatively low...
%
% ^ <hannah.buckley@aut.ac.nz> 2017-08-30T00:26:24.142Z.
\end{abstract}

%-------------------------------------------------------------------------------------------------
% INTRODUCTION
%-------------------------------------------------------------------------------------------------
\section{Introduction}\label{sec:intro}

Spatial associations are a fundamental aspect of most ecological and environmental data. Although accounting for spatial covariation has become routine in ecological data analysis \citep{Fortin:2005}, ecologists and environmental scientists have been slower to appreciate and account for anisotropic patterns and processes \citep[but see, e.g.,][] {Ellison:2014}. Codispersion \citep{Vallejos:2015} measures lag-dependent spatial covariation in two or more spatial processes, which may be anisotropic. Codispersion recently has been used to analyze ecological data and provide new insights into potential ecological processes that underlie observed patterns in co-occurrence between pairs of species \citep{Buckley:2016a} and in relationships between attributes of individual species and the underlying environment \citep{Buckley:2016b}.

Applications of codispersion analysis that have been published to date have assumed either that there are no errors in the datasets or that any errors that are present would have no effect on the analysis. These assumptions are clearly unrealistic. The goal of this paper is to better understand how sensitive the codispersion coefficient is to different types of noise and measurement error ("contamination") in the analyzed data. We approach this goal by using Monte Carlo simulation studies to examine several classes of noise that we would expect to occur in datasets or images analyzed using codispersion.

We consider the effect on codispersion of simple observation error, in which noise is added to a fixed number of random points (or pixels) in a dataset (or image) either as white noise (spatially independent and identically distributed) or as a spatially-dependent process. We also consider images with missing values distributed either randomly throughout the sample space or in clusters (e.g., clouds obscuring a large section of a remotely-sensed image) and present different algorithms for interpolation prior to calculation of codispersion. Using bivariate data for forest tree-environment relationships, we evaluate the robustness of codispersion under different levels of error introduced into the environmental data (kriged surfaces derived from complete or "thinned" datasets mimicking those with missing values). In most cases, the effect of these sources of noise and error are tested in one of two ways: either by calculating the codispersion between the original dataset (image) and a contaminated version of itself (codispersion values are correlated to quantify the effect of the noise) or, in the case of the species-environment relationships, by calculating the codispersion between the two original datasets and then comparing the output to the codispersion calculated between the tree data and contaminated environmental data.

In Section \ref{Methods}, we describe the different types of contamination and observation error that we added to both real and simulated datasets. We also describe the method we used for imputing missing data. Results are presented in Section \ref{Results}, and discussed in Section \ref{Discussion}. Technical details on simulation and imputation algorithms are given in the Appendix. 

%------------------------------------------------------------------------------------------------
% Methods
%------------------------------------------------------------------------------------------------
\section{Methods}\label{Methods}
In spatial modeling and time series, several types of contamination can be specified \citep{Anselin:1995, Fox:1972}. Here, we consider types of contamination frequently observed in spatial data. Many of our examples are from data collected on forest trees. These data will be familiar to many ecologists and environmental scientists, and to date, codispersion analysis used in ecological settings has been applied primarily to forested ecosystems.

\begin{enumerate}
\item[\bf 1.] {\bf Salt-and-pepper noise on an image}: Salt-and-pepper noise has been used widely in image processing and computational statistics to represent real distortions \citep{Huang:2010} and to generate different scenarios via Monte Carlo simulation \citep{McQuarrie:2003}.

Assume that $X$ is an image and suppose that $X$ follows a zero mean normal distribution with variance $\sigma^2.$ Then we consider additive noise following a zero mean normal distribution with variance $\tau^2$ such that $\tau^2 \gg \sigma^2.$ The contamination is located randomly in space such that a small percentage of observations are corrupted with a probability $\delta.$ Specifically, 
\begin{equation}\label{cont1}
X\sim (1-\delta)\mathcal{N}(0,\sigma^2)+\delta \mathcal{N}(0,\tau^2).
\end{equation}
Preliminary experiments were introduced by \cite{Vallejos:2015} but here we present a more extensive study. The contamination scheme was generated by using Monte Carlo simulations according to \eqref{cont1}. We considered $\sigma^2=1$, $\tau^2=1, 5, 10,$ and the percentage of contamination $\delta=0.05, 0.1, 0.25.$ We conjecture that the codispersion coefficient is robust for $\delta\leq 0.05.$
% \begin{figure}[h!]
% \centering
% \subfloat[]{\includegraphics[width=6.0cm, height=4.0cm]{gray_image2.png}}
% \hspace{2mm}
% \subfloat[]{\includegraphics[width=7.0cm, height=4.6cm]{Image3D_GI2.jpg}}\\
% \subfloat[]{\includegraphics[width=6.0cm, height=4.0cm]{SPG2image005.png}}
% \hspace{2mm}
% \subfloat[]{\includegraphics[width=7.0cm, height=4.6cm]{SP3D_GI2005.jpg}}\\
% \subfloat[]{\includegraphics[width=6.0cm, height=4.0cm]{SPG2image015.png}}
% \hspace{2mm}
% \subfloat[]{\includegraphics[width=7.0cm, height=4.6cm]{SP3D_GI2015.jpg}}\\
% \subfloat[]{\includegraphics[width=6.0cm, height=4.0cm]{SPG2image025.png}}
% \hspace{2mm}
% \subfloat[]{\includegraphics[width=7.0cm, height=4.6cm]{SP3D_GI2025.jpg}}
% \caption{\label{fig:gray_image2} (a) Reference image of size $5616 \times 3744$ pixels taken above a section of forest at the Harvard Forest, Petersham, MA and (b) its corresponding gray scale values. (c),(e) and (g) are the same image distorted with salt-and-pepper noise. The percentages of contamination are 5\%, 10\%, and 25\% respectively. (d), (f), and (h) show the change in gray intensity after the addition of salt-and-pepper noise to the images.}
% \end{figure}

\begin{figure}[h!!]
\centering
\includegraphics[scale=1]{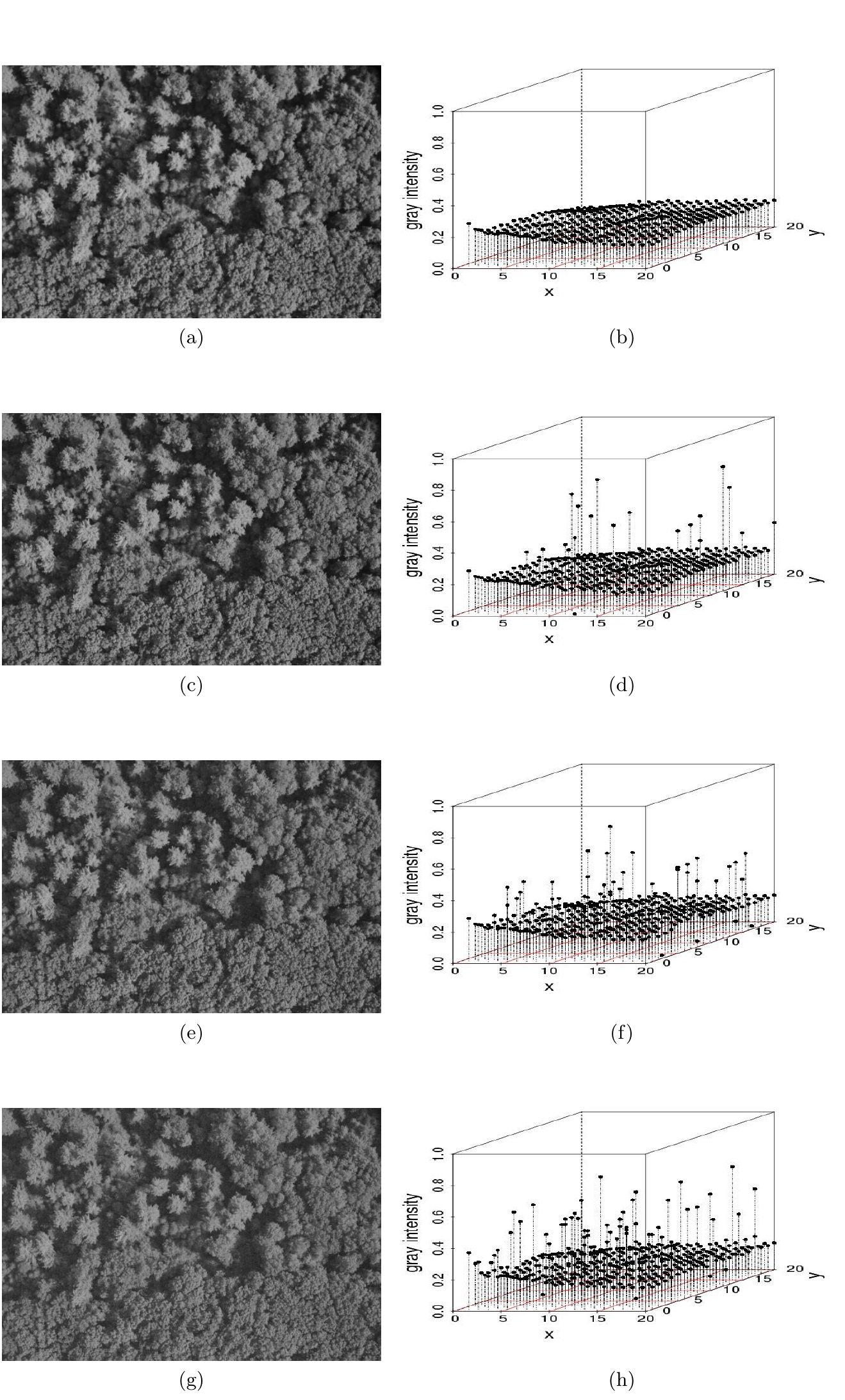}
\caption{\label{fig:gray_image2} (a) Reference image of size $5616 \times 3744$ pixels taken above a section of forest at the Harvard Forest, Petersham, MA and (b) its corresponding gray scale values. (c),(e) and (g) are the same image distorted with salt-and-pepper noise. The percentages of contamination are 5\%, 10\%, and 25\% respectively. (d), (f), and (h) show the change in gray intensity after the addition of salt-and-pepper noise to the images.}
\end{figure}

In Figure \ref{fig:gray_image2}(a) we illustrate an aerial photograph of a forest stand in Massachusetts, USA. Figures \ref{fig:gray_image2} (c),(e), and (g) are contaminated versions of the original one when $\delta=\{0.05, 0.10, 0.25\}$. The corresponding perspective plots shown in Figure \ref{fig:gray_image2}(b),(d),(f), and (h) depict the effect of contamination on the gray intensities. The greater the contamination, the greater the dispersion as is observed in the z-axis of the three dimensional scatter plots displayed in Figure \ref{fig:gray_image2}.
 
We compared codispersion calculated for the original image to that calculated for the contaminated images. In addition to the reference image shown in Figure \ref{fig:gray_image2}(a) we considered other aerial images. Figure \ref{fig:image_reference} displays four images with different textures related to the same forest stand in Massachusetts. The codispersion maps of these images are presented in the supplementary material for this paper.

\begin{figure}[h!]
\centering
\subfloat[]{\includegraphics[width=6cm, height=4cm]{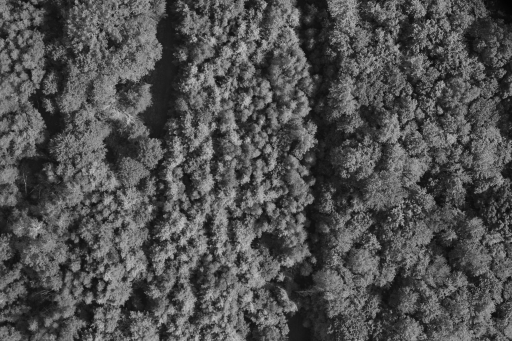}}
\hspace{2mm}
\subfloat[]{\includegraphics[width=6cm, height=4cm]{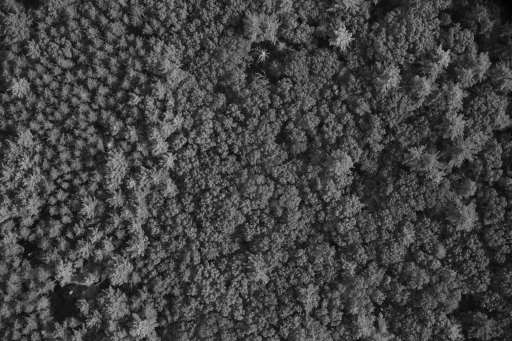}}\\
\subfloat[]{\includegraphics[width=6cm, height=4cm]{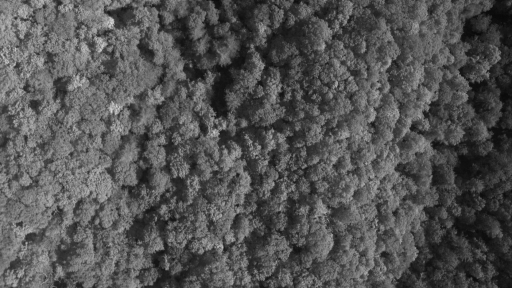}}
\hspace{2mm}
\subfloat[]{\includegraphics[width=6cm, height=4cm]{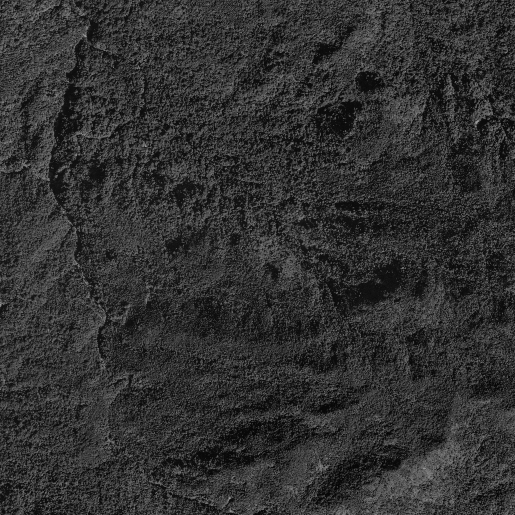}}
\caption{\label{fig:image_reference} Images (a)-(d) represent other kind of textures that can be found in the 
forest stand in Massachusetts from aerial photographs.}
\end{figure}

\item[\bf 2.] {\bf Salt-and-pepper on Dependent Processes}: 
\cite{Gneiting:2010} extended the well known Mat\'ern class of covariance functions to a multivariate random field. In particular, a bivariate spatial process $(X(\bm s), Y(\bm s))^{\top}$, where $\bm s\in D\subset \mathbb{R}^2$, has a Gaussian distribution with mean vector zero and a matrix-valued covariance function
\begin{equation}\label{eq:bivariate}
\left(
\begin{matrix}
C_{11}(\bm h) & C_{12}(\bm h)\\
C_{21}(\bm h) &C_{22}(\bm h)
\end{matrix}
\right),
\end{equation}
where $\bm h \in D,$ $C_{ii}(\bm h)=\sigma_iM(\bm h|a_i)$ for $i=1,2$, $C_{12}=C_{21}=\rho_{12}\sigma_1\sigma_2M(\bm h|\nu_{12},a_{12})$, 
$$M(\bm h| \nu, a)=\frac{2^{1-\nu}}{\Gamma(\nu)}(a||\bm h||)^{\nu}K_{\nu}(a||\bm h||)$$ is the correlation at lag distance $\bm h$ with $K_{\nu}$ (a modified Bessel function of the second kind), and $a>0.$ The parsimonious bivariate Matérn model has the restriction 
$$|\rho_{12}|\leq \frac{(\nu_1\nu_2)^{1/2}}{\frac{1}{2}(\nu_1+\nu_2)}.$$ The correlation between the spatial variables $X(\cdot)$ and $Y(\cdot)$ is controlled by the parameter $\rho_{12},$ which allows one to generate bivariate Gaussian spatial processes with different levels of dependence. Without loss of generality, it can be assumed that the mean of the bivariate process is zero, but the theory works well for any bivariate process with mean $(\mu_1,\mu_2)^{\top}.$ Any type of contamination can be applied over the generated dependence data. In this case, we applied salt-and-pepper noise.

We generated dependent random fields from the bivariate Mat\'ern class of covariance functions described in Equation \eqref{eq:bivariate} by Monte Carlo simulation using the R package RandomFields \citep{Schlather:2017}. We then added the salt-and-pepper noise, varying the additional parameter $\rho_{12}$, which represents the known correlation between processes $X(\cdot)$ and $Y(\cdot)$.

\begin{figure}[h!!]
\centering
\includegraphics[scale=1.05]{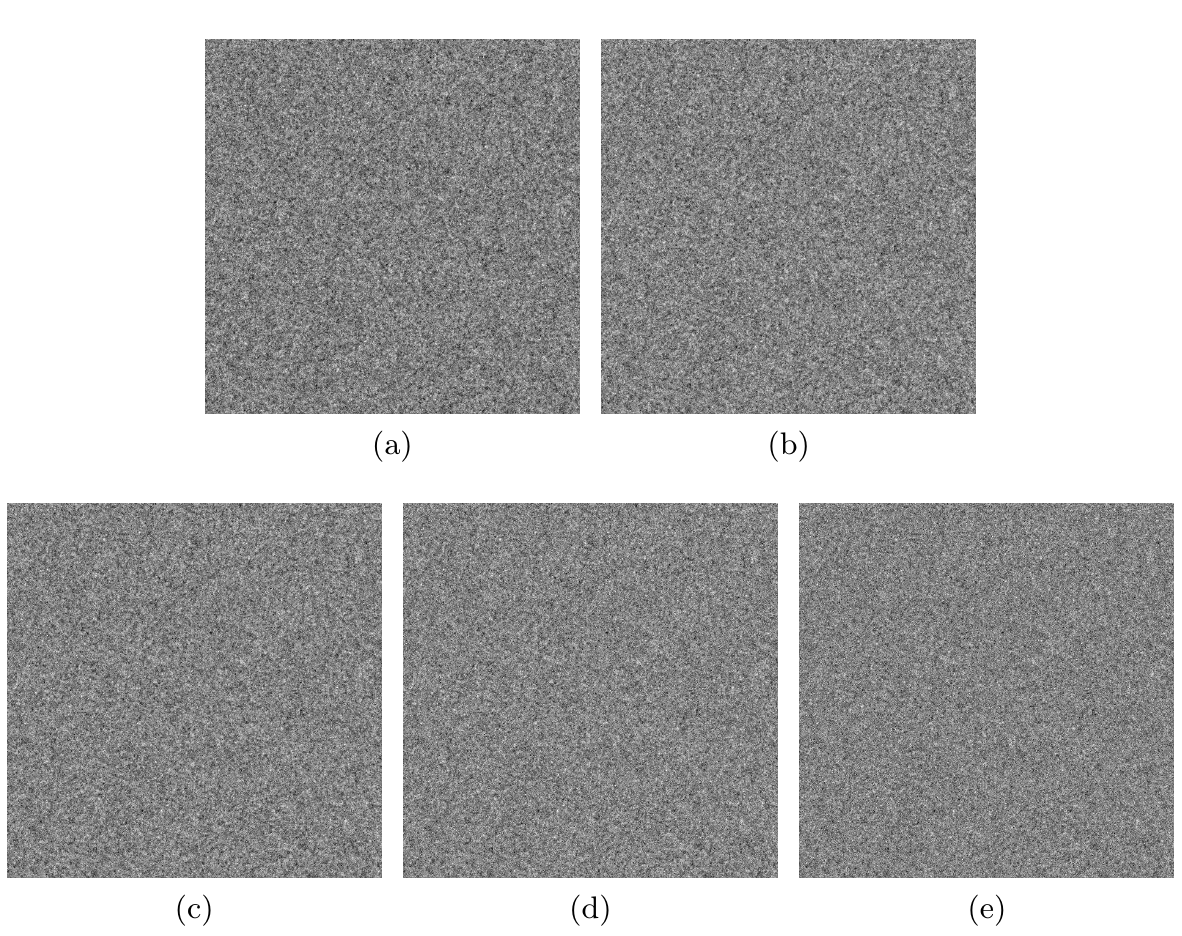}
\caption{\label{fig:dependent1} Images (a) and (b) are dependent processes generated from a Gaussian process with a covariance matrix as in \eqref{eq:bivariate}; (c)-(e) salt and pepper contamination of (b) considering $\delta=(0.05,0.15,0.25)$ and $ (\nu_1, \nu_2, \sigma^2_1, \sigma^2_2, \mu_1, \mu_2, \rho_{12})=( 0.5,0.5,1,1,0.5,0.5,0.8)$ in each case.}
\end{figure}

Figure \ref{fig:dependent1} shows one realization of size $512\times 512$ from a bivariate Gaussian process with correlation equal to 0.8 (images (a) and (b)). Figure \ref{fig:dependent1} (c),(d), and (e) show versions of (b) contaminated with salt-and-pepper noise with the percentage of contamination equal to 5\%, 15\%, and 25\%, respectively. Because the Gaussian process is stationary, images (a) and (b) look very smooth and regular, and any correlation between them (if it exists) is difficult to observe in the printed images.

\item[\bf 3.] {\bf Missing Observations at Random Locations}: We used the salt-and-pepper scheme to intentionally delete \textit{n} observations at random. We first defined the percentage of contamination ($\delta$), and then deleted that many observations from the dataset. In practice we replaced observations with NA at the randomly-selected locations. The main feature of these missing observations is that they are spatially independent of one another, but for the posterior data analysis they will remain fixed. The imputation Algorithm described in the Appendix was not applied here because codispersion calculations are not affected when the percentage of contamination $\delta$ is small. 

% \begin{figure}[h!!]
% \centering
% \subfloat[]{\includegraphics[width=4.845cm, height=3.23cm]{MORL_GI2031.png}}
% \hspace{1mm}
% \subfloat[]{\includegraphics[width=4.845cm, height=3.23cm]{MORL_GI2032.png}}
% \hspace{1mm}
% \subfloat[]{\includegraphics[width=4.845cm, height=3.23cm]{MORL_GI2033.png}}\\
% \subfloat[]{\includegraphics[width=4.845cm, height=3.23cm]{MORL_GI2021.png}}
% \hspace{1mm}
% \subfloat[]{\includegraphics[width=4.845cm, height=3.23cm]{MORL_GI2022.png}}
% \hspace{1mm}
% \subfloat[]{\includegraphics[width=4.845cm, height=3.23cm]{MORL_GI2023.png}}\\
% \subfloat[]{\includegraphics[width=4.845cm, height=3.23cm]{MORL_GI2011.png}}
% \hspace{1mm}
% \subfloat[]{\includegraphics[width=4.845cm, height=3.23cm]{MORL_GI2012.png}}
% \hspace{1mm}
% \subfloat[]{\includegraphics[width=4.845cm, height=3.23cm]{MORL_GI2013.png}}
% \caption{\label{fig:missing2} Contamination of of the reference image shown in Figure \ref{fig:gray_image2}(a) by salt and pepper at random locations. The missing blocks are of size $15\times15$, $30\times30$, and $60\times60$ respectively, shown in the different columns. (a)-(c): The proportion of missing blocks is 0.000002. (d)-(f): The proportion of missing blocks is 0.000004.
% (d)-(f): The proportion of missing blocks is 0.000008.}
% \end{figure}

\begin{figure}[h!!]
\centering
\includegraphics[scale=1]{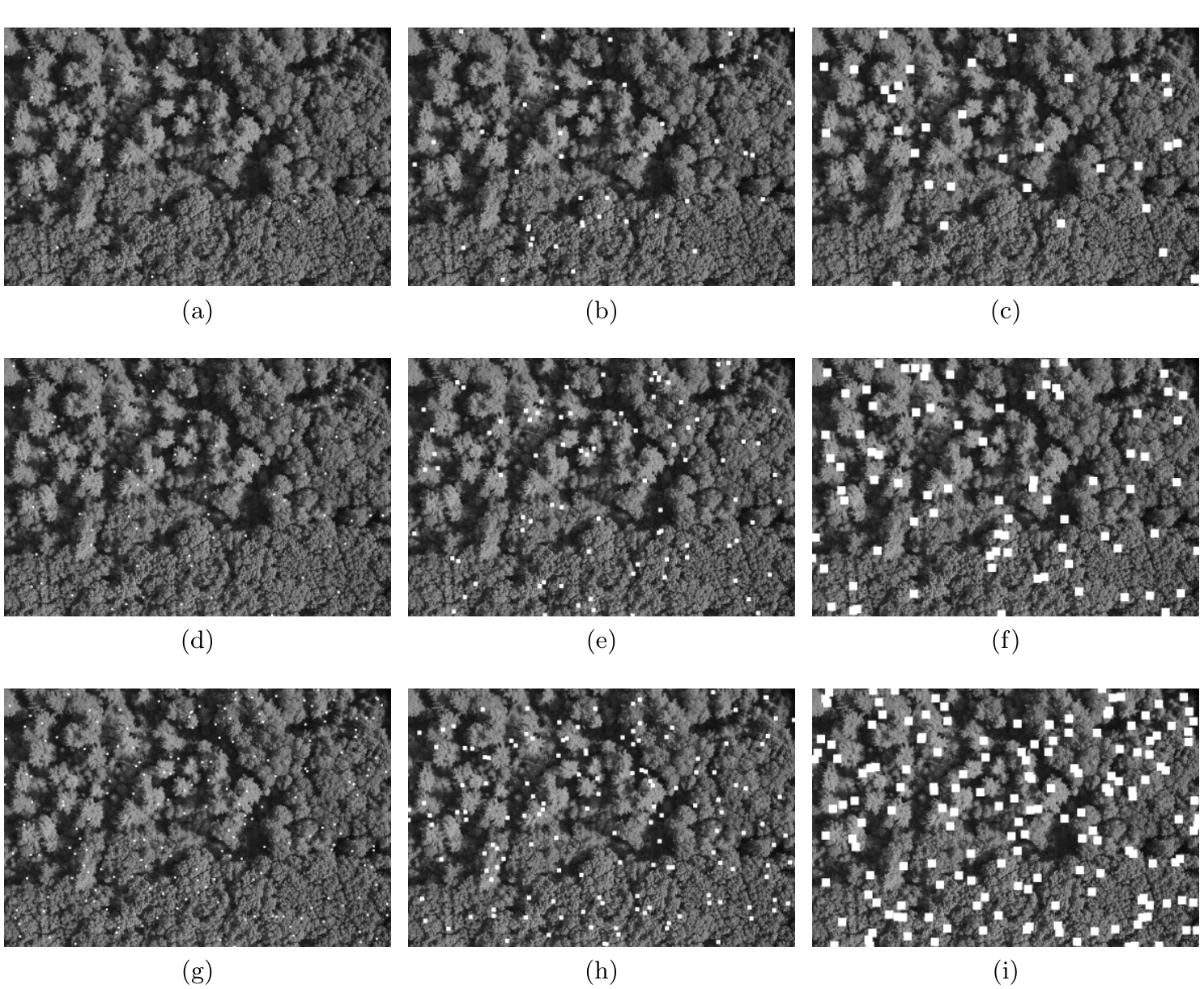}
\caption{\label{fig:missing} Contamination of of the reference image shown in Figure \ref{fig:gray_image2}(a) by salt and pepper at random locations. The missing blocks are of size $15\times15$, $30\times30$, and $60\times60$ respectively, shown in the different columns. (a)-(c): The proportion of missing blocks is 0.000002. (d)-(f): The proportion of missing blocks is 0.000004.
(d)-(f): The proportion of missing blocks is 0.000008.}
\end{figure}

In Figure \ref{fig:missing} we illustrate the missing random scheme with nine contaminated versions of the original image shown in Figure \ref{fig:gray_image2}(a). The columns show the effect of increasing the percentage of contamination (5\%, 15\%, and 25\% respectively), and the rows depict the effect of increasing the block size of contaminated pixels, which are $15\times15$, $30\times30$, and $60\times60$ respectively. The contaminated pixels have been colored in white. NAs were ignored in the computation of the codispersion coefficients because for large gaps of missing observations the computation of the codispersion coefficient will be affected for those directions $\bm h$ such that $||\bm h||$ is less than the maximum diameter of the missing block. 

\item[\bf 4.] {\bf Gaps Resulting from Clusters of Missing Observations}: Missing values may be clustered, for example, either because of local difficulties in sampling or because large sections of an image are obscured by clouds or shadows. We simulated clustered missing observations for the image shown in Figure \ref{fig:gray_image2}(a), given three different pixel sizes for the contaminated block: $200\times 200$, $400\times 400$, and $800\times 800$ (Figure \ref{fig:missing_gaps}). We used simple clustered geometries (squares) for ease of computation. The difference between the previous type of contamination and this one is that in the former, the contamination consisted of several blocks of small size. Here we introduced just one gap containing a large number of pixels which, in Figure \ref{fig:missing_gaps}, is located for illustrative purposes in the center of the image. However, in our simulations and analysis, we randomized both the size of the missing block and its location. 
% \begin{figure}[htp]
% \centering
% \subfloat[]{\includegraphics[width=4.845cm, height=3.23cm]{MORL3_GI2011.png}}
% \hspace{1mm}
% \subfloat[]{\includegraphics[width=4.845cm, height=3.23cm]{MORL3_GI2012.png}}
% \hspace{1mm}
% \subfloat[]{\includegraphics[width=4.845cm, height=3.23cm]{MORL3_GI2013.png}}
% \caption{\label{fig:missing_gaps1} Gaps of missing observations of sizes $200 \times 200$ (a); $400\times 400$ (b); and $800\times 800$.}
% \end{figure}

\begin{figure}[h!!]
\centering
\includegraphics[scale=1]{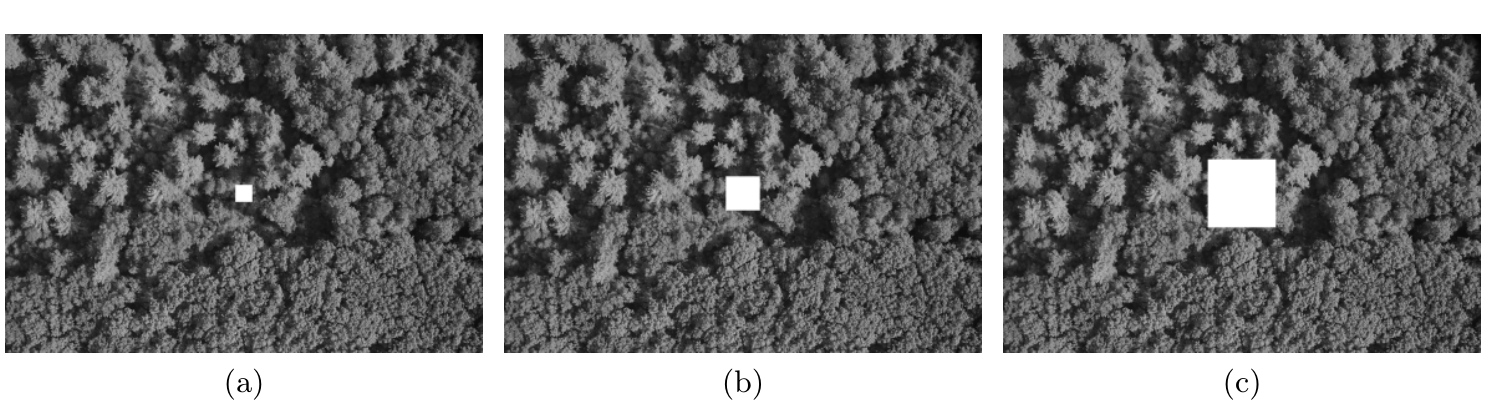}
\caption{\label{fig:missing_gaps} Gaps of missing observations of sizes $200 \times 200$ (a); $400\times 400$ (b); and $800\times 800$.}
\end{figure}

To compute codispersion coefficients for datasets with such large blocks of missing data, we needed to fill the missing gaps (impute missing data) prior to computing the codispersion coefficient. We address this issue in two ways. 

First, the image with a missing gap is represented by a first-order spatial autoregressive process. The fitting of the parameters of the models is done via least squares estimation following the guidelines given by \cite{Allende:2001}. This estimation method was studied by \cite{Ojeda:2010} and found to yield an approximated image $\widehat{Z}$ of the original one $X$ (see Algorithm \ref{alg:alg1} in the Appendix).

Second, to predict the values of the process in the locations belonging to the missing block, a reconstruction is made by applying Algorithm \ref{alg:alg1} to the four closest blocks to the missing gap as is illustrated in Figure \ref{fig:prediction}. This prediction scheme is summarized in Algorithm \ref{alg:pred} (Appendix). In simple words, the first step represents the image intensity by an autoregressive process that assumes that the intensity of any pixel is a weighted average of the intensity of the surrounding pixels. This is a model-based alternative to the average or median commonly computed using the intensities of a moving window across the image. The second step predicts the missing values using similar autoregressive models to represent the surrounding blocks. The predicted value of a pixel belonging to the missing block is a weighted average where the weights are proportional to the distance from the missing pixel to the surrounding blocks

\item[\bf 5.] {\bf Sampling Error}: Spatial data often are sampled from a kriged surface, which itself is generated from a set of field observations. The information in the kriged surface is a function of the number of observations and the smoothing parameter of the covariance function \citep{Minasny:2005}. For a pair of spatial point processes $X(\cdot)$ and $Y(\cdot)$, where the number of observations in $X(\cdot)$ and $Y(\cdot)$ differed by several orders of magnitude, we generated a kriged surface from $Y(\cdot)$ using either all or two random ("thinned") subsets of forest plot data (tree species locations and soil chemistry variables), containing respectively 90\% and 80\% of the original soil chemistry data \citep{Minasny:2005}. We then sampled the kriged surfaces to obtain predicted values $\hat{Y}(\cdot)$ for each of the sampled points (tree locations) in $X(\cdot)$. 

To assess sampling error, we used data from plants and soils collected in the 50-ha forest dynamics plot on Barro Colorado Island, Panam\'a \citep{Condit:1998, Hubbell:1998, Hubbell:2005}. Of the 299 plant species mapped, identified, and measured every five years in this plot, we used six: \emph{Alseis blackiana, Oenocarpus mapora, Hirtella triandra, Protium tenuifolium, Poulsenia armata, and Guarea guidonia} (Figure \ref{fig:BCIVeg}). The abundances of unique single-stemmed individuals of each of these six species ranged from 993 (\emph{Poulsenia armata}) to 7928 (\emph{Alseis blackiana}), and included species that had a range of positive, negative, and weak associations with measured soil variables \citep{John:2007}. Spatial locations and diameters of individual trees of each species (excluding dead individuals and individuals with more than one stem) were taken from the seventh (2010) semi-decadal census of the plot. 

\begin{figure}[h!]
\centering
\includegraphics[scale=0.48]{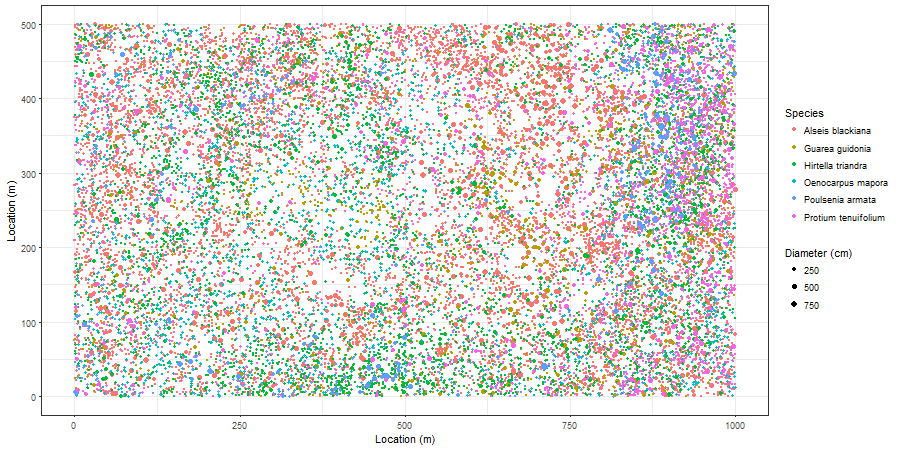}
\caption{Distribution and size of the six species of trees growing in the 50-hectare plot at Barro Colorado Island, Panam\'a that we analyzed to assess the effect of sampling error.}
\label{fig:BCIVeg}
\end{figure}

Soil samples were collected on a 50-m lattice in 2005 with additional samples taken at finer spatial grains at alternate sampling stations \citep{John:2007}. Soil samples were analyzed for concentrations of 11 elements; we used only data for concentrations of calcium (Ca), phosphorus (P), and aluminium (Al), which had the highest loadings on the first three principal axes of a multivariate analysis (NMDS) on the complete soil dataset \citep{John:2007}. We used ordinary kriging in the geoR package \citep{Ribeiro:2001}, version 1.7-5.2, to fit a surface to the data for each soil element and predict its concentration at the location of each tree (Figure \ref{fig:BCISoils}). Variogram models (exponential, exponential, and wave for Ca, P, and Al, respectively) needed as input for the kriging function were fit to detrended (2\textsuperscript{nd}-order polynomial) data that had been Box-Cox transformed ($\lambda$ = 0.5, 1.0, and 1.0 for Ca, P, and Al, respectively); kriging was done on back-transformed data to which the trend had been added. Nuggets were estimated empirically for Ca and P, but the nugget for Al was fixed (following visual inspection of the empirical variogram) equal to 4,000. 

\begin{figure}[h!]
\centering
\includegraphics[scale=0.75]{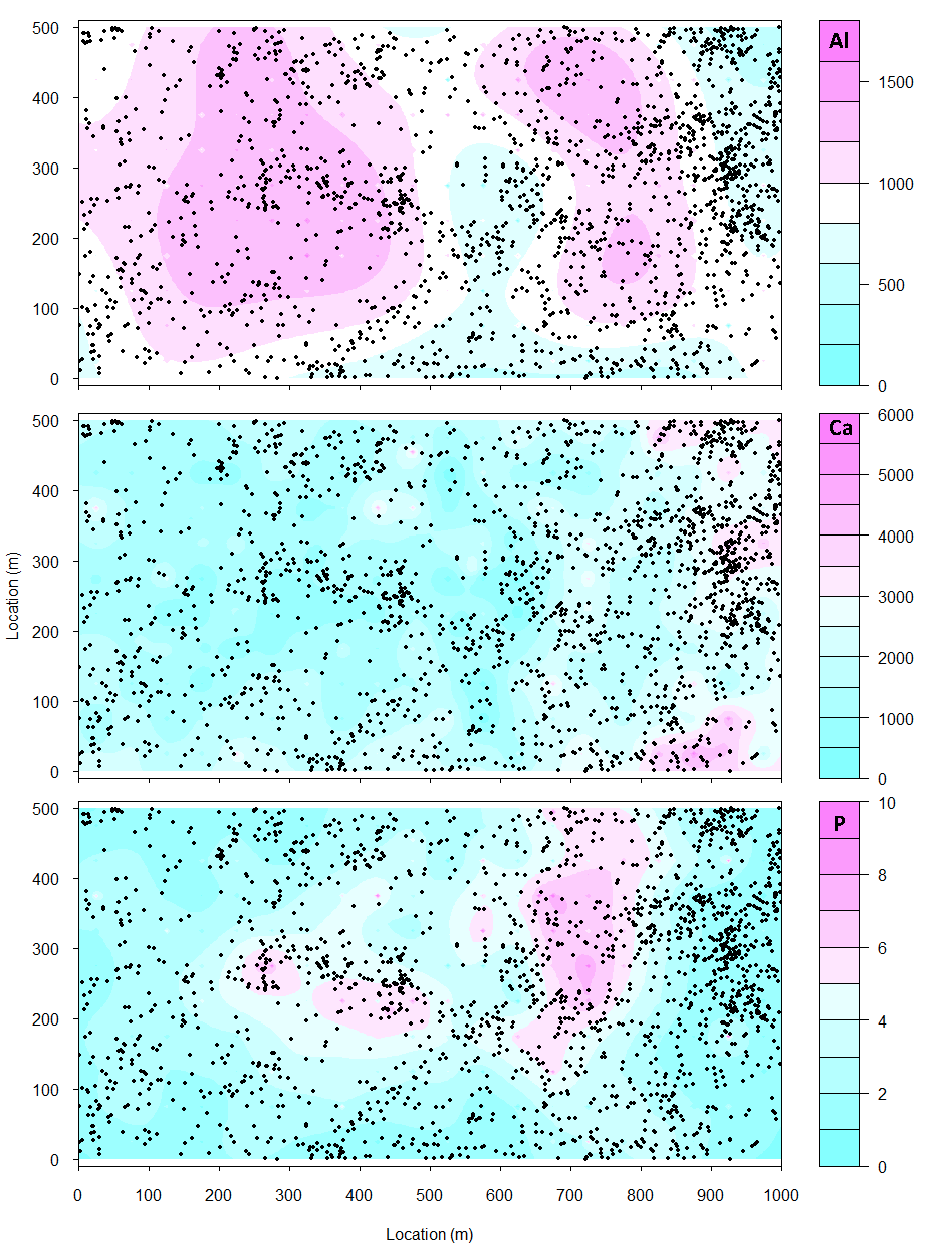}
\caption{Kriged surfaces of the concentration (mg/kg) of aluminum (Al; top), calcium (Ca; center), and phosphorus (P; bottom) in the 50-hectare plot at Barro Colorado Island, Panam\'a. Contours were estimated for a regular grid (5-m spacing) based on data from samples taken at the individual points shown on the plots.}
\label{fig:BCISoils}
\end{figure}
\end{enumerate}

\subsection{Availability of data and code}

BCI vegetation and soils data are available from 
\url{http://ctfs.si.edu/webatlas/datasets/bci/}. Analyses were done using the R software system, version 3.3.1 \citep{R:2016}. The images and all the code used in this paper are available from (\url{blinded/  website})%\url{https://github.com/JAcostaS/rcode}.

%------------------------------------------------------------------------------------------------
% Results
%------------------------------------------------------------------------------------------------
\section{Results}\label{Results}
We used codispersion maps \citep{Buckley:2016a, Vallejos:2015} to explore the possible patterns and features caused by the introduction of noise and to evaluate the performance of the codispersion coefficient when the process was contaminated with one of the distortions described above. Recall that the generation of the noise is through statistical models that do not necessarily include a particular direction in space. The effects of specific directional contamination on codispersion was investigated by \cite{Vallejos:2015}.

\begin{figure}[!htp]
\centering
\subfloat[]{\includegraphics[width=5.2cm, height=3.7cm]{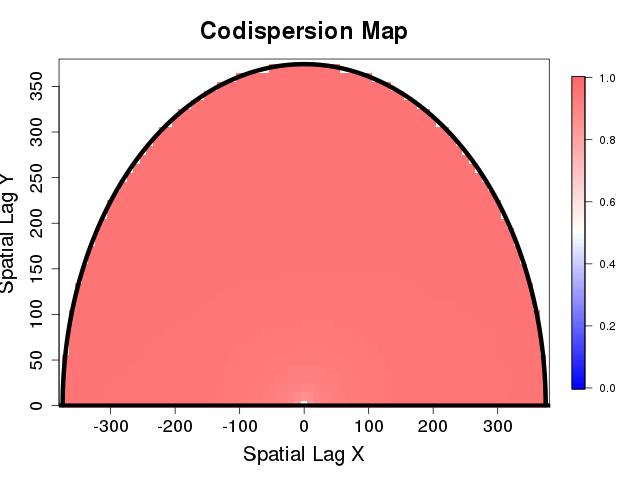}}
\subfloat[]{\includegraphics[width=5.2cm, height=3.7cm]{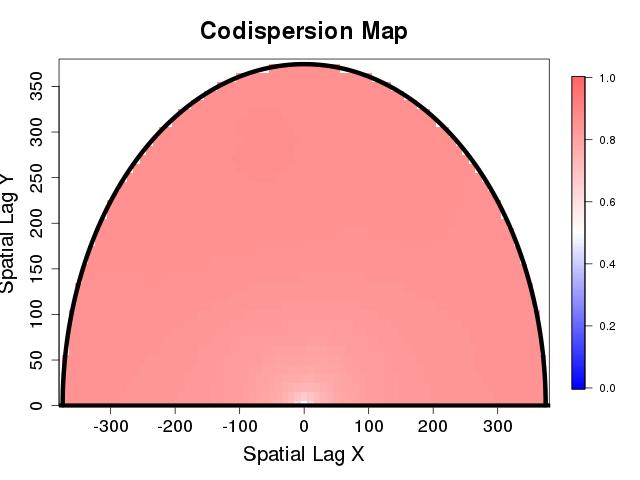}}
\subfloat[]{\includegraphics[width=5.2cm, height=3.7cm]{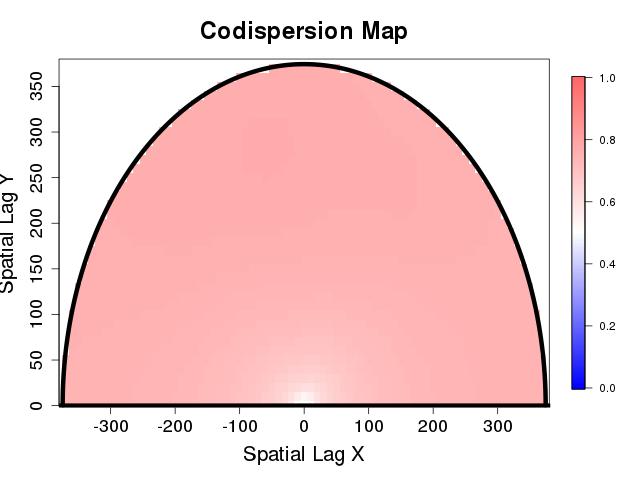}}
\caption{\label{fig:salt and pepper} 
Codispersion map between the original image \ref{fig:gray_image2}(a) and images 
\ref{fig:gray_image2} (c), (e) and (g) (5\%, 15\% and 25\% respectively) contaminated with salt and pepper noise.}
\end{figure}

\begin{figure}[h!!]
\centering
\subfloat[]{\includegraphics[width=5.2cm, height=3.7cm]{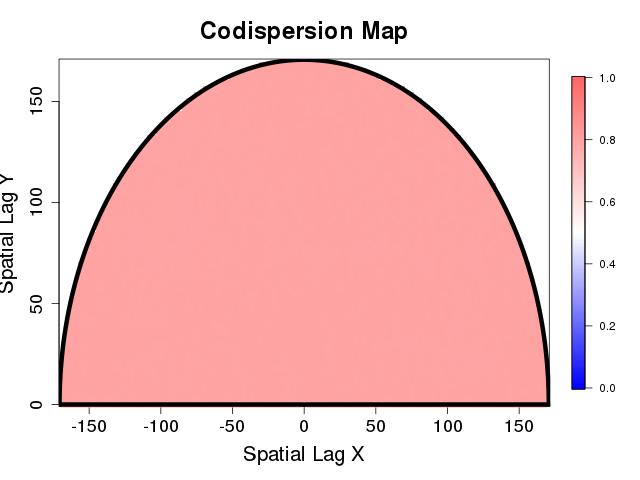}}
\subfloat[]{\includegraphics[width=5.2cm, height=3.7cm]{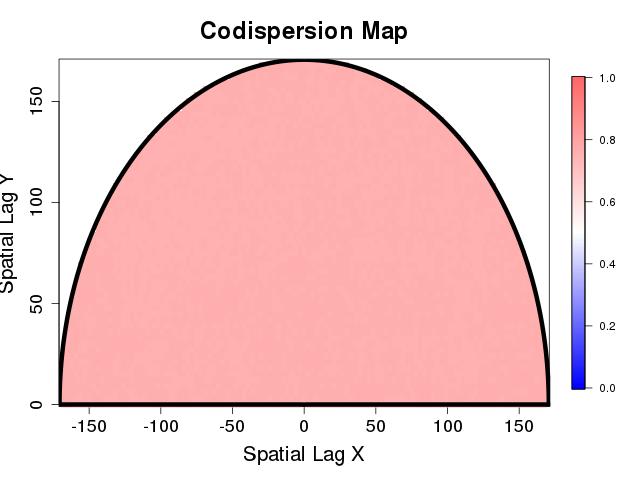}} \\
\subfloat[]{\includegraphics[width=5.2cm, height=3.7cm]{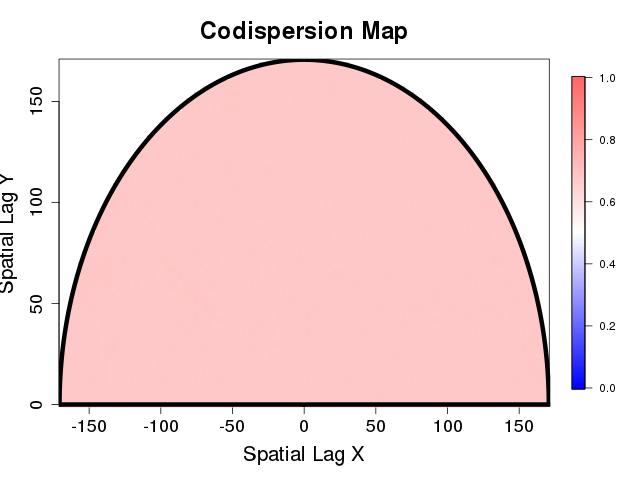}}
\subfloat[]{\includegraphics[width=5.2cm, height=3.7cm]{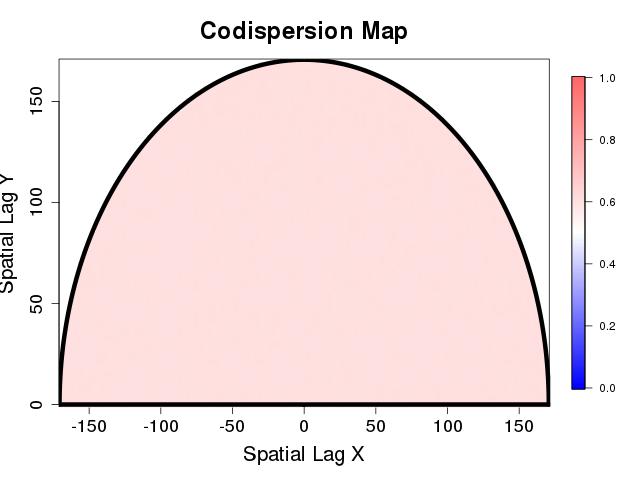}} 
\caption{\label{fig:dependent2} Images (a)-(d) 
are the corresponding codispersion maps between image \ref{fig:dependent1}(a) and the contaminated images (b)-(e).}
\end{figure}

\begin{figure}[!htp]
\centering
\subfloat[]{\includegraphics[width=5.2cm, height=3.7cm]{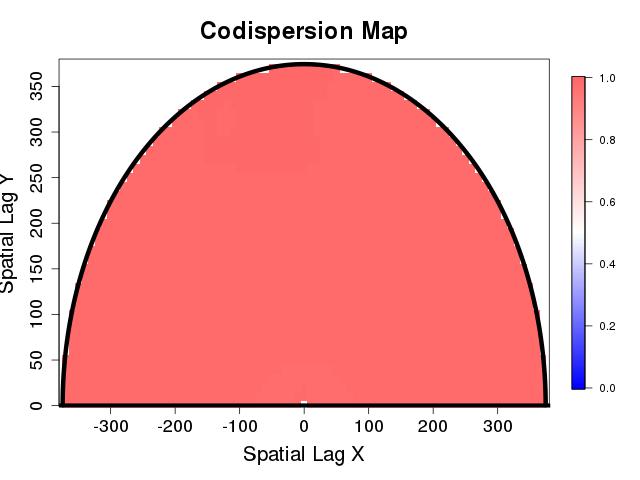}}
\subfloat[]{\includegraphics[width=5.2cm, height=3.7cm]{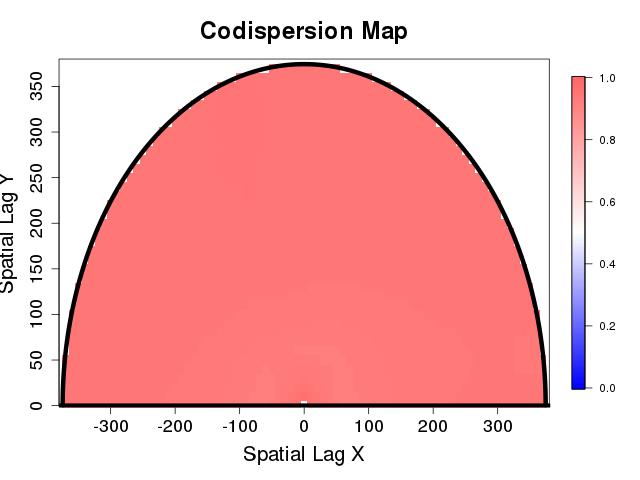}}
\subfloat[]{\includegraphics[width=5.2cm, height=3.7cm]{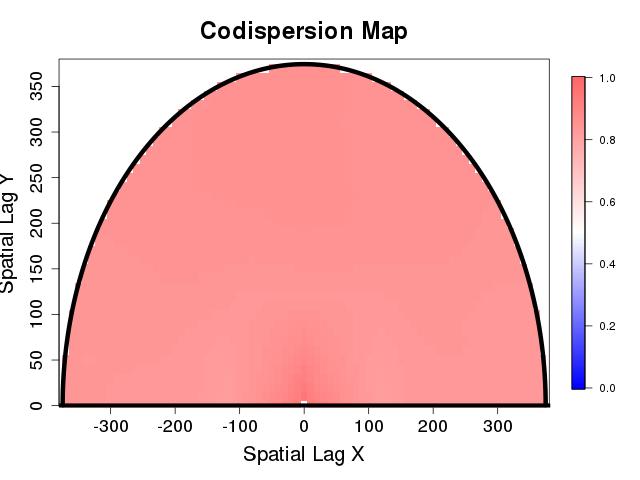}} \\
\subfloat[]{\includegraphics[width=5.2cm, height=3.7cm]{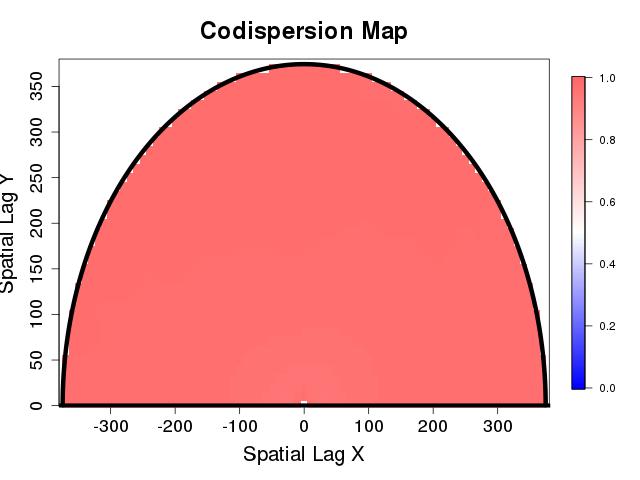}}
\subfloat[]{\includegraphics[width=5.2cm, height=3.7cm]{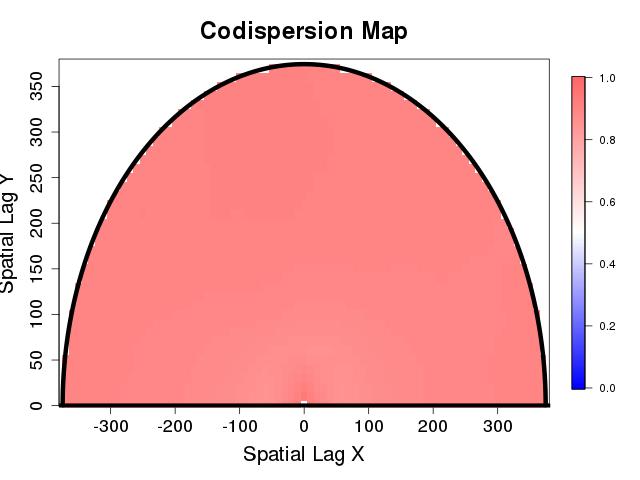}}
\subfloat[]{\includegraphics[width=5.2cm, height=3.7cm]{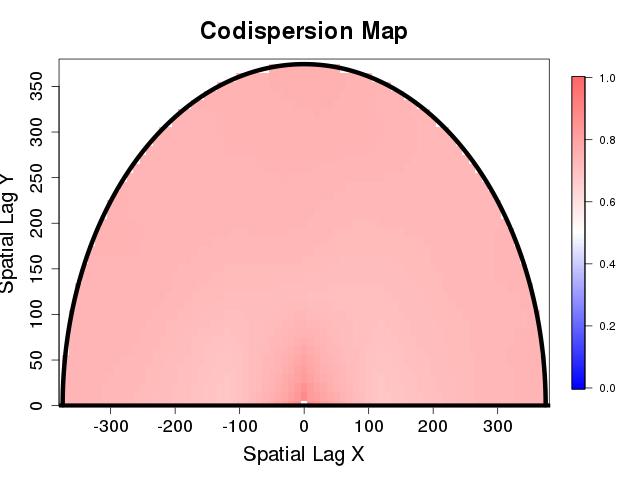}} \\
\subfloat[]{\includegraphics[width=5.2cm, height=3.7cm]{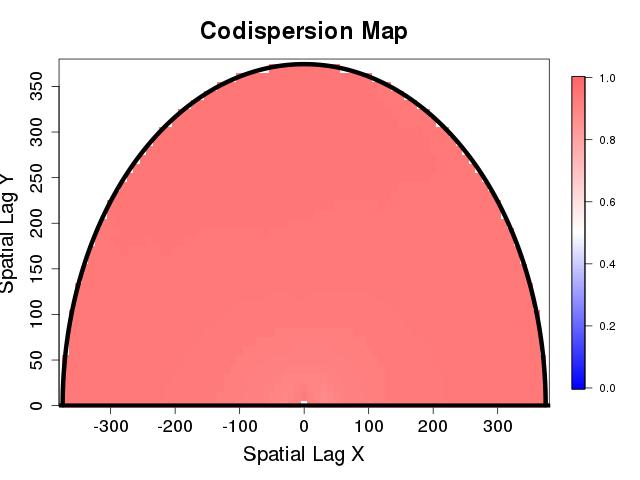}}
\subfloat[]{\includegraphics[width=5.2cm, height=3.7cm]{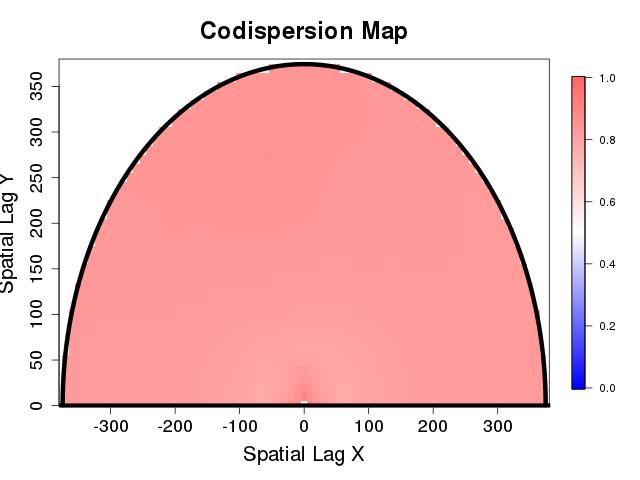}}
\subfloat[]{\includegraphics[width=5.2cm, height=3.7cm]{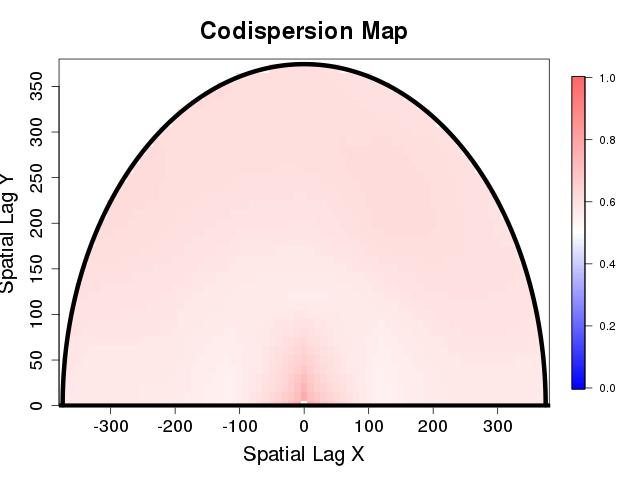}} 
\caption{\label{fig:Missing Observation} Codispersion map between the reference image \ref{fig:gray_image2}(a) and the images contaminates with missing observation at random locations depicted in Figure \ref{fig:missing}(a-i).}
\end{figure}

The only effect observed when the forest image was contaminated with salt-and-pepper noise (Figure \ref{fig:gray_image2}) was a trend of decreasing codispersion between the original and contaminated images with an increase in the percentage of contamination (Figure \ref{fig:salt and pepper}). In the case of the dependent processes generated by a Gaussian process with covariance matrix as in \eqref{eq:bivariate}, we plotted the codispersion maps between the original and contaminated images displayed in Figure \ref{fig:dependent1}. The salt-and-pepper contamination caused a complete loss of correlation between the two images, which were originally correlated by 0.8 (Figure \ref{fig:dependent2}). A decrease in codispersion between the original and contaminated images was also observed when noise was introduced through missing observations at random locations or as the missing block size increased (Figure \ref{fig:Missing Observation}).

\begin{figure}[h!!]
\centering
\subfloat[]{\includegraphics[width=4.845cm, height=3.23cm]{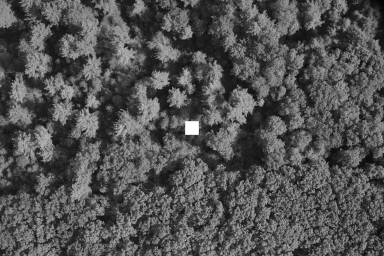}}
\hspace{1mm}
\subfloat[]{\includegraphics[width=4.845cm, height=3.23cm]{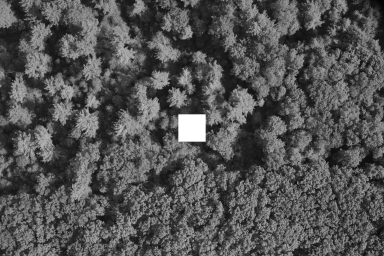}}
\hspace{1mm}
\subfloat[]{\includegraphics[width=4.845cm, height=3.23cm]{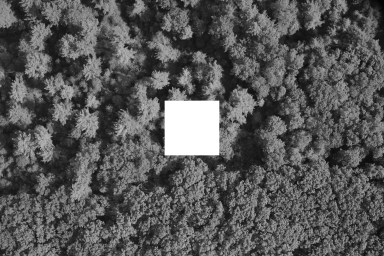}} \\
\subfloat[]{\includegraphics[width=4.845cm, height=3.23cm]{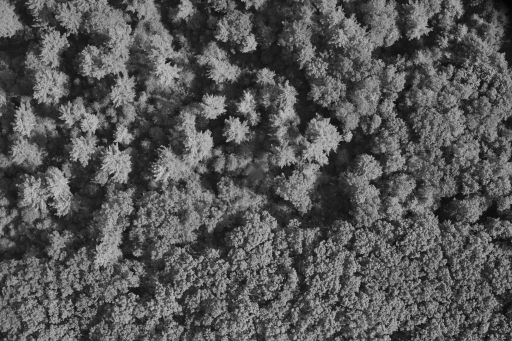}}
\hspace{1mm}
\subfloat[]{\includegraphics[width=4.845cm, height=3.23cm]{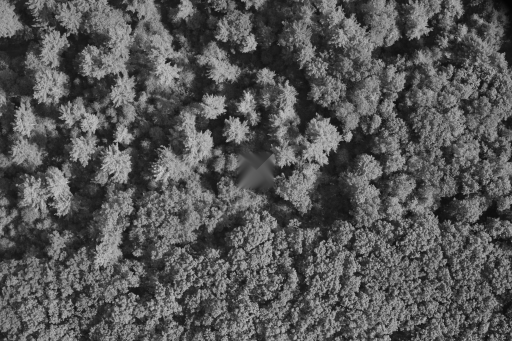}}
\hspace{1mm}
\subfloat[]{\includegraphics[width=4.845cm, height=3.23cm]{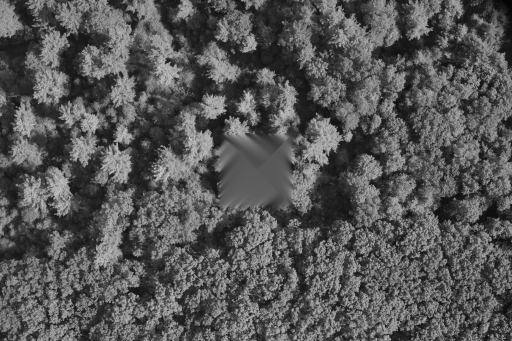}} \\
\subfloat[]{\includegraphics[width=5.2cm, height=3.7cm]{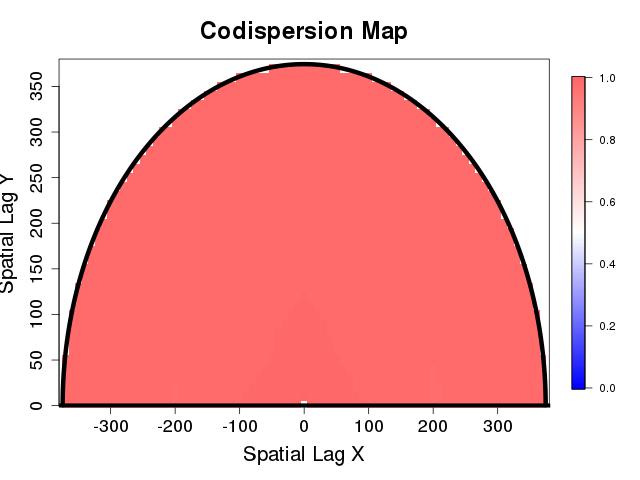}}
\subfloat[]{\includegraphics[width=5.2cm, height=3.7cm]{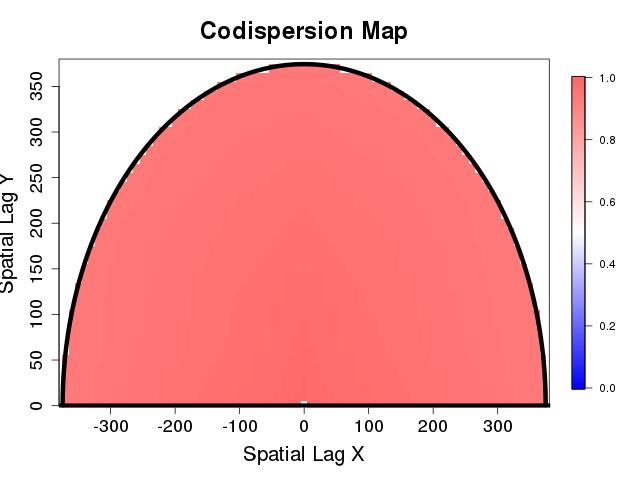}}
\subfloat[]{\includegraphics[width=5.2cm, height=3.7cm]{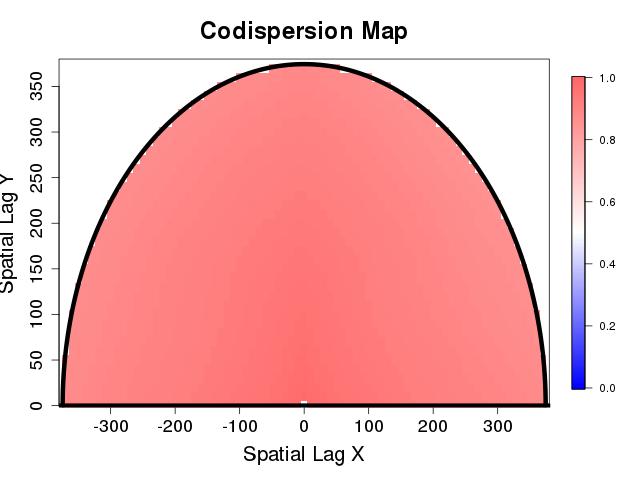}} \\
\subfloat[]{\includegraphics[width=5.2cm, height=3.7cm]{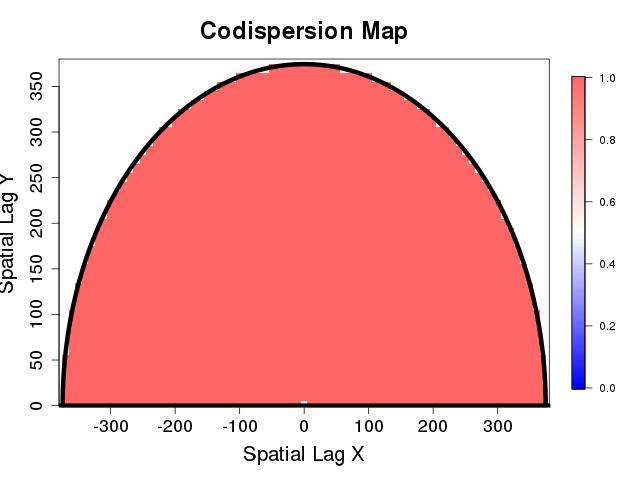}}
\subfloat[]{\includegraphics[width=5.2cm, height=3.7cm]{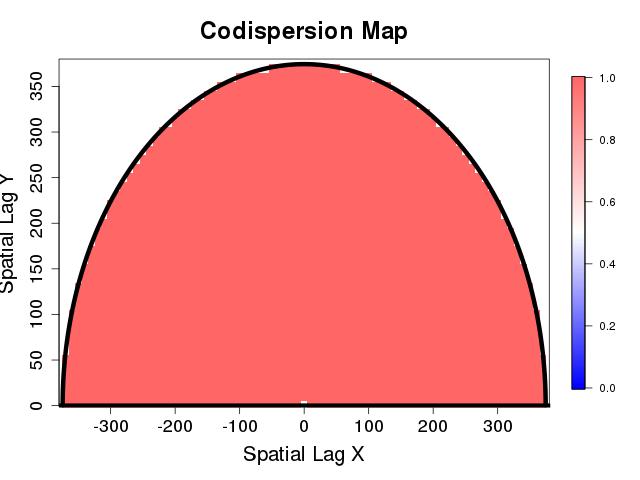}}
\subfloat[]{\includegraphics[width=5.2cm, height=3.7cm]{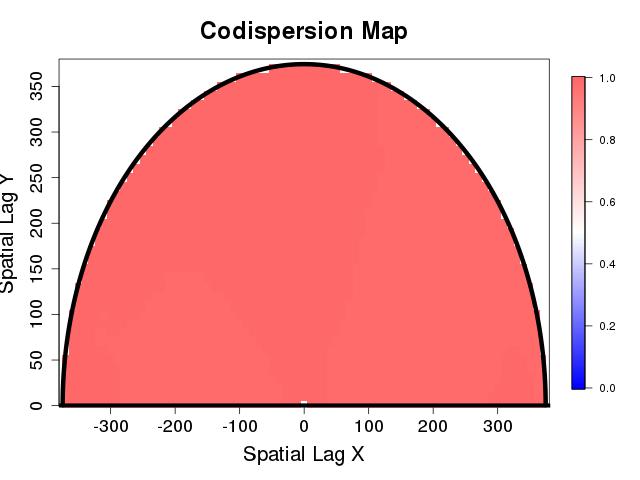}} 

\caption{\label{fig:gap missing} 
Contamination of the reference image \ref{fig:gray_image2}(a)
by gaps resulting from clusters of mussing observations. Images (a)-(c) contain only one missing block in the center of the image of sizes $200\times200$, $400\times400$, y $800\times800$ respectively. Images (d)-(f) were yielded by the imputation algorithm described in the Appendix. Images (g)-(i) are the corresponding codispersion maps between image \ref{fig:gray_image2}(a) and the imputed images (d)-(f).}
\end{figure}

Figure \ref{fig:missing_gaps} illustrates how we introduced large gaps of missing values in the center of the reference image shown in Figure \ref{fig:gray_image2}(a). Before computing the codispersion map, we imputed the missing data (Algorithm \ref{alg:pred} in the Appendix). Although the performance of such algorithms strongly depends on the size of the block of missing observations, the construction of it is based on the spatial information contained in the nearest neighbors (Algorithm \ref{alg:alg1} in the Appendix). The spatial autoregressive lags in the AR-2D process are fixed when the order of the process is chosen. In this case, three neighbors were considered in a strongly causal set to guarantee an infinite moving average representation of the process. The images filled by the imputation algorithm are shown in Figure \ref{fig:gap missing}(d)-(f). The filled areas are smooth in terms of texture and have a smaller variance. Visually, the imputation of the larger missing block looks different from the rest of the image. For small missing blocks it is difficult to see the imputed values. From Figure \ref{fig:gap missing}(g)-(h), we observed that Algorithm \ref{alg:pred} was able to recover valuable information and that the codispersion between the original and imputed images in all cases was close to one.

\begin{figure}[h!!]
\centering
\subfloat[Al]{\includegraphics[width=5.32cm, height=3.99cm]{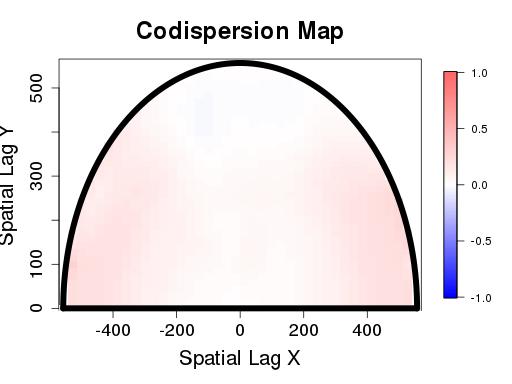}}
\subfloat[Al90]{\includegraphics[width=5.32cm, height=3.99cm]{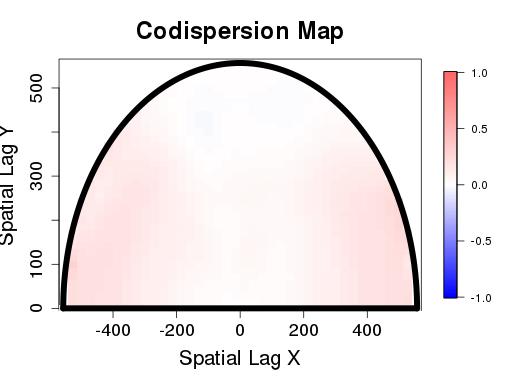}}
\subfloat[Al80]{\includegraphics[width=5.32cm, height=3.99cm]{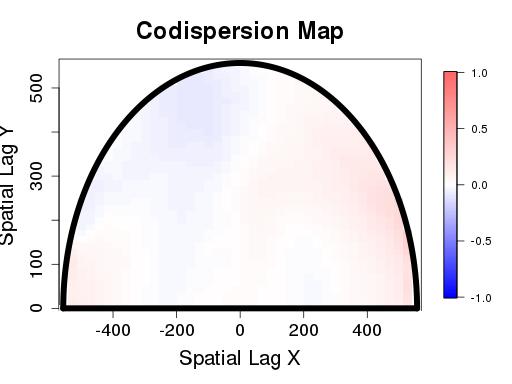}}\\
\subfloat[Ca]{\includegraphics[width=5.32cm, height=3.99cm]{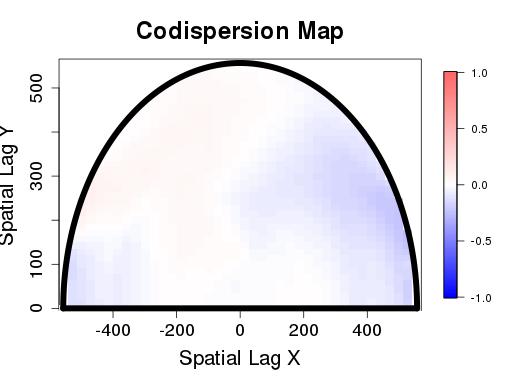}}
\subfloat[Ca90]{\includegraphics[width=5.32cm, height=3.99cm]{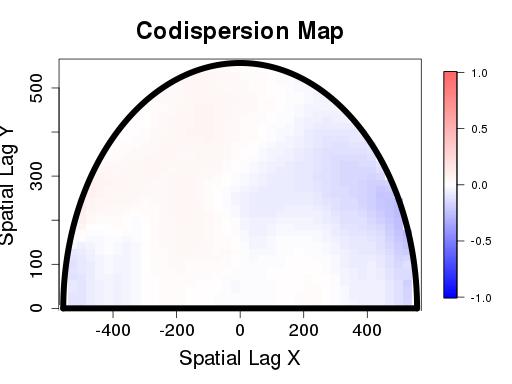}}
\subfloat[Ca80]{\includegraphics[width=5.32cm, height=3.99cm]{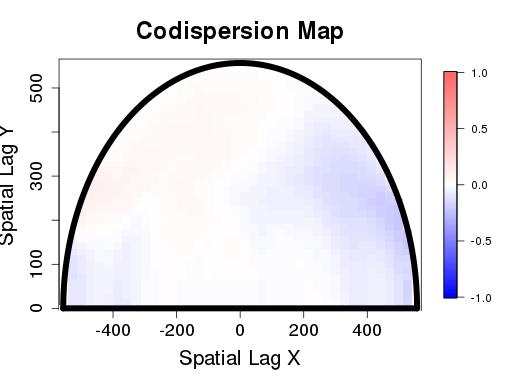}}\\
\subfloat[P]{\includegraphics[width=5.32cm, height=3.99cm]{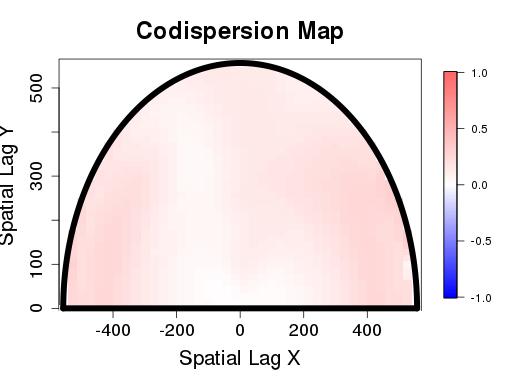}}
\subfloat[P90]{\includegraphics[width=5.32cm, height=3.99cm]{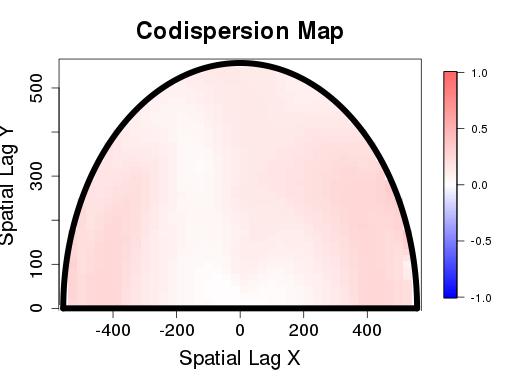}}
\subfloat[P80]{\includegraphics[width=5.32cm, height=3.99cm]{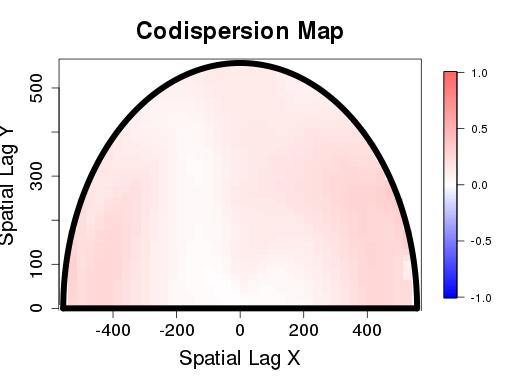}}\\
 \caption{\label{fig:specie1} Codispersion between species \textit{A. blackiana} and soil chemistry variables; (Al (a)-(c); Ca (d)-(f) and P (g)-(i)). Soils data were unthinned (a, d, g), thinned 10\% (b, e, h), or thinnd 20\% (c, f, i).}
\end{figure}

\begin{figure}[h!!]
\centering
\subfloat[Al]{\includegraphics[width=5.32cm, height=3.99cm]{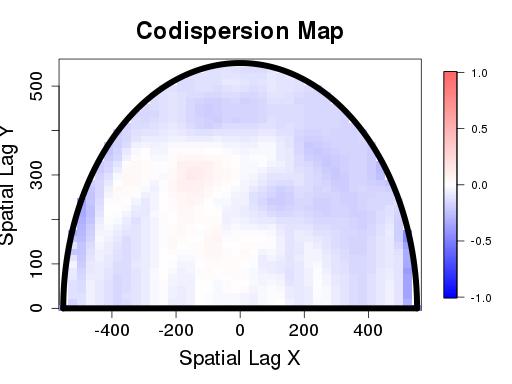}}
\subfloat[Al90]{\includegraphics[width=5.32cm, height=3.99cm]{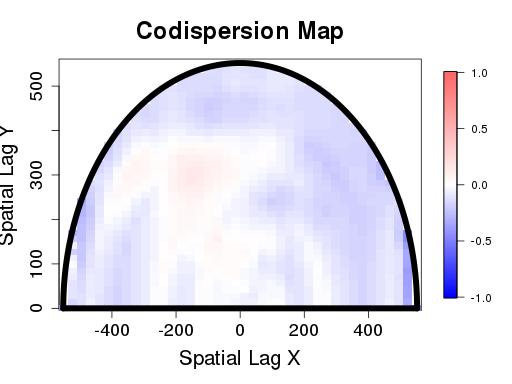}}
\subfloat[Al80]{\includegraphics[width=5.32cm, height=3.99cm]{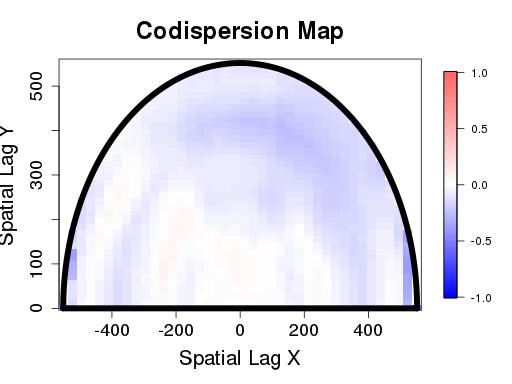}}\\
\subfloat[Ca]{\includegraphics[width=5.32cm, height=3.99cm]{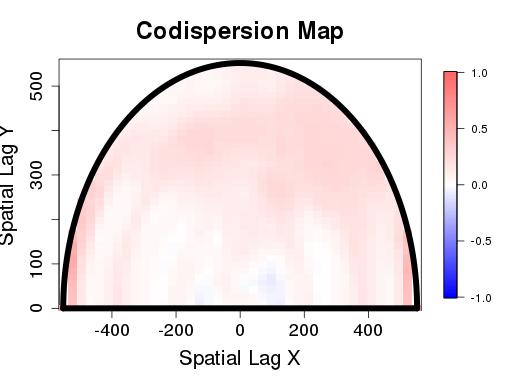}}
\subfloat[Ca90]{\includegraphics[width=5.32cm, height=3.99cm]{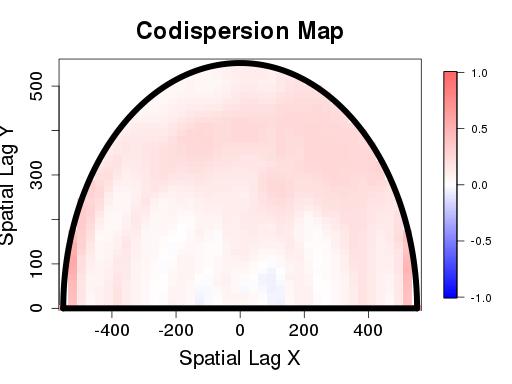}}
\subfloat[Ca80]{\includegraphics[width=5.32cm, height=3.99cm]{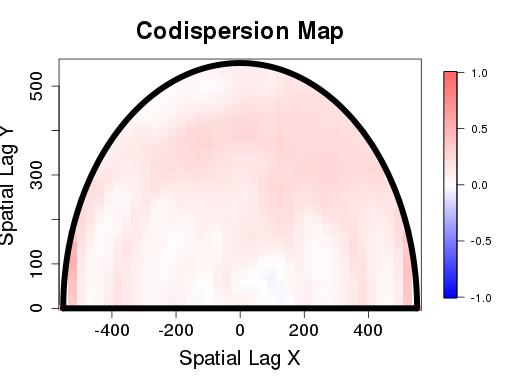}}\\
\subfloat[P]{\includegraphics[width=5.32cm, height=3.99cm]{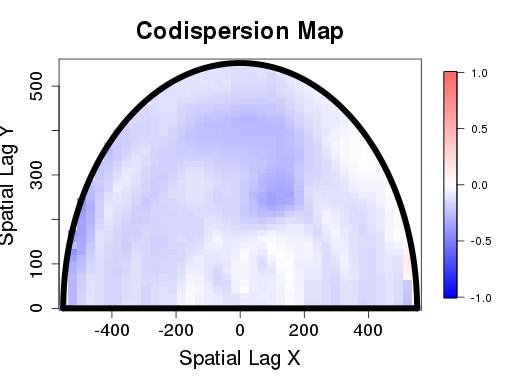}}
\subfloat[P90]{\includegraphics[width=5.32cm, height=3.99cm]{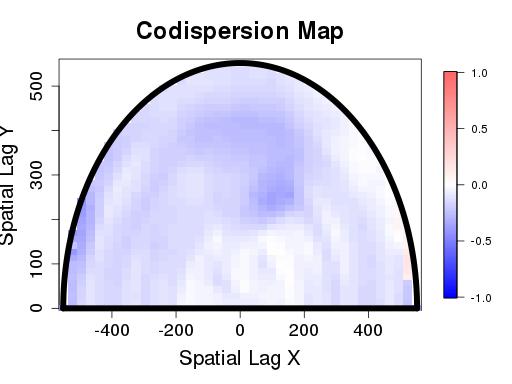}}
\subfloat[P80]{\includegraphics[width=5.32cm, height=3.99cm]{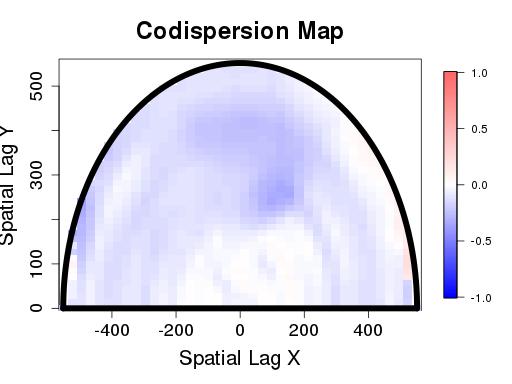}}\\
\caption{\label{fig:specie5} Codispersion between species \textit{P. armata} and soil chemistry variables; (Al (a)-(c); Ca (d)-(f) and P (g)-(i)). Soils data were unthinned (a, d, g), thinned 10\% (b, e, h), or thinnd 20\% (c, f, i). }
\end{figure}

Finally, the codispersion between tree species' diameters (for the six species shown in Figure \ref{fig:BCIVeg}) and the three soil elements (Figure \ref{fig:BCISoils}) sampled in the Barro Colorado Island plot at three levels of data 'thinning' (none, 10\% and 20\%) showed that codispersion was robust to this form of contamination. Only the results for the most abundant (Figure \ref{fig:specie1}) and the least abundant (Figure \ref{fig:specie5}) species are shown.

%------------------------------------------------------------------------------------------------
% DISCUSSION
%------------------------------------------------------------------------------------------------
\section{Discussion}\label{Discussion}
The methods and examples developed in this paper improve our understanding of the behavior of the codispersion coefficient when data have been contaminated. The codispersion coefficient appears to be robust for small percentages of contamination (< 15\%), but always leads to an underestimation of the codispersion between the datasets. As the percentage of contamination increases, the codispersion decreases in all directions on the plane. We note that the types of noise considered in this paper did not affect the codispersion in any particular direction(s). Although the performance of codispersion for directional noise was explored by \cite{Vallejos:2015,Vallejos:2016}, directtional noise has not yet been observed in real datasets.

When applied to environmental data, codispersion has been shown to useful for describing scales of covariation in two or more variables across complex spatial gradients (e.g., Buckley et al. 2016a, b). Our ability to detect such spatial pattern depends on the grain of spatial variation in the data and how this compares to the lag sizes used in the codispersion analysis. For example, the complete loss of correlation between the two images in Figure \ref{fig:dependent1} under only a small degree of contamination highlights the importance of considering the spatial grain of the datasets relative to that of the noise-inducing processes. The coarser-grained spatial pattern in the forest images is retained, even under contamination, whereas the spatial dependence in the images in Figure \ref{fig:dependent1} is at a smaller grain than the extent of the image, which is relatively heavily disturbed by the salt-and-pepper noise. 

The imputation algorithm described in the Appendix seems to be a promising technique to handle blocks of missing observations. Several aspects of it are worth exploring with future research. These include the success of the algorithm in recovering missing observations as a function of the block size; how to select the number of neighbors to be considered in the AR-2D process; and the similarity between the texture of the imputed observations and the texture of the reference image. For simplicity and without loss of generality, the missing blocks we illustrated were square regions located in the center of the image, but certainly Algorithm \ref{alg:pred} could be extended to other types of regions located anywhere in the image. 

More general aspects of codispersion analysis are in need of further exploration and testing. First, it will be of interest to study the results of codispersion analysis of rasterized images. This is because rasterization of images is widespread and common rasterization methods rarely, if ever, preserve the original spatial correlation of each process. The development of a new rasterization method that preserves better the spatial correlation within processes could follow \cite{Goovaerts:2010}. Second, the computation of codispersion maps is computationally expensive. Thus, the development of efficient algorithms capable of creating codispersion maps for large images is still needed.
%------------------------------------------------------------------------------------------------
% ACKNOWLEDGEMENTS
%------------------------------------------------------------------------------------------------
\section*{Acknowledgements}
%RV was partially supported by AC3E, FB-0008, Chile. JA was supported by a Chilean National Graduate Scholarship. AME's participation in this project was supported by Harvard University and the Universidad T\'ecnica Federico Santa Mar\'ia, and HLB's and BSC's work on this project in Chile also was supported by the Universidad T\'ecnica Federico Santa Mar\'ia. Vegetation data from Barro Colorado Island (BCI) are part of the BCI forest dynamics research project founded by S. P. Hubbell and R. B. Foster and now managed by R. Condit, S. Lao, and R. Perez through the Center for Tropical Forest Science (CTFS) and the Smithsonian Tropical Research Institute (STRI) in Panam\'a. Numerous organizations have provided funding to support this long-term study, principally the US National Science Foundation, and hundreds of field workers have contributed to mapping, measuring and monitoring the vegetation. Jim Dalling, Robert John, Kyle Harms, Robert Stallard, Joe Yavitt, Paolo Segre, and Juan Di Trani sampled the soils at BCI. Collection and initial analysis of the BCI data were supported by NSF grants 021104, 021115, 0212284, 0212818 and 0314581, the STRI Soils Initiative, and CTFS. This paper is a publication of the Harvard Forest Long-Term Ecological Research Site, supported by the US National Science Foundation. 

%------------------------------------------------------------------------------------------------
% Appendix
%------------------------------------------------------------------------------------------------
\bigskip
\bigskip
\appendix
 \noindent {\bf \Large Appendix}
\numberwithin{equation}{section}

%-------------------------------------------------------------------------------------------------
% Appendix
%-------------------------------------------------------------------------------------------------
\section{Image Imputation Algorithm}
The algorithm described below is based on the fact that it is possible to represent any image by using unilateral AR-2D processes \citep{Ojeda:2010}. The generated image is called a local AR-2D approximated image by using blocks.

Let $Z=\{ Z_{r,s}: 0\leq r\leq M-1,0\leq s\leq N-1 \}$ be an original image, and let
$X$ the original image corrected by the mean. That is, $X_{r,s}=Z_{r,s}-\overline{Z},$
for all $0\leq r\leq M-1,$ $0\leq s\leq N-1,$ and for which $\overline{Z}$ is the mean of $Z$.

Following \cite{Bustos:2009}, assume that $X$ follows a causal AR-2D process of the form 
\begin{equation*}
X_{r,s}=\phi _{1}X_{r-1,s}+\phi_{2}X_{r,s-1}+\phi_{3}X_{r-1,s-1}+\varepsilon_{r,s},
\end{equation*}%
where $(r,s)\in \mathbb{Z}^{2}$, $\left( \varepsilon_{r,s} \right) _{(r,s)\in \mathbb{Z}^{2}}$ is Gaussian white noise, and $\phi_1,\phi_2$, and $\phi_3$ are the autoregressive parameters.

Let $4\leq k\leq \min (M,N).$ For simplicity we consider that the images to be processed are
arranged in such a way that the number of columns minus one and the number of rows minus one are multiples of $k-1$; Then we define the $(k-1)\times (k-1)$ block $\left(i_{b},j_{b}\right) $ of the image $X$ by%
\begin{equation*}
B_{X}\left( i_{b},j_{b}\right) =\{ X_{r,s}: (k-1)(i_{b}-1)+1\leq r\leq (k-1)i_{b},(k-1)(j_{b}-1)+1\leq s\leq (k-1)j_{b}\},
\end{equation*}
for all $i_{b}=1,\cdots ,\left[ (M-1)/(k-1)\right] $ and for all $ j_{b}=1,\cdots ,\left[ (N-1)/(k-1)\right]$, where $[\cdot]$ denotes the integer part. The $M^{\prime }\times N^{\prime }$ approximated image $\widehat{Z}$, where $M^{\prime } =\left[(M-1)/(k-1)\right] (k-1)+1$ and $N^{\prime }=\left[ (N-1)/(k-1)\right] (k-1)+1$ can be obtained by the following algorithm.

\begin{algorithm}[h!!]
\caption{Approximated AR-2D Image.}\label{alg:alg1}
\label{alg}
{\bf Input:} An original image $Z$ of size $M\times N$.

{\bf Output:} An approximated $\widehat{Z}$ of size $M^{\prime}\times N^{\prime}.$ 
\begin{algorithmic}[1]
\State {\bf for each block $B_{X}\left( i_{b},j_{b}\right)$ do}
\State \hspace{3mm} Compute the least square (LS) estimators of $\phi _{1}$, $\phi _{2}$ and $\phi_3$ associated with block $B_{X}\left( i_{b},j_{b}\right).$ 
\State \hspace{3mm} Define $\widehat{X}$ on the block $B_{X}\left( i_{b},j_{b}\right) $ by
\begin{equation*}
\widehat{X}_{r,s}=\widehat{\phi }_{1}\left( i_{b},j_{b}\right) X_{r-1,s}+%
\widehat{\phi }_{2}\left( i_{b},j_{b}\right) X_{r,s-1}+\widehat{\phi }_{3}\left( i_{b},j_{b}\right) X_{r-1,s-1},
\end{equation*}%
\hspace{3mm} where $(k-1)(i_{b}-1)+1\leq r\leq (k-1)i_{b}$, $(k-1)(j_{b}-1)+1\leq s\leq (k-1)j_{b},$ and $\widehat{\phi }_{1}\left( i_{b},j_{b}\right)$, 

\hspace{-4mm} $\widehat{\phi }_{2}\left( i_{b},j_{b}\right)$, and $\widehat{\phi }_{3}\left( i_{b},j_{b}\right)$
are the LS estimators of $\phi_1, \phi_2$ and $\phi_3$ respectively.
\State {\bf end for}
\State \hspace{3mm} The approximated image $\widehat{Z}$ of $Z$ is:
\begin{equation*}
\widehat{Z}_{r,s}=\widehat{X}_{r,s}+\overline{Z},\text{ \ \ }0\leq
r\leq M^{\prime }-1,0\leq s\leq N^{\prime }-1.
\end{equation*}
\State {\bf Return} $\widehat{Z}$.
\end{algorithmic}
\end{algorithm}

\begin{figure}[h!!]
 \centering
 \includegraphics[width=6cm, height=6cm]{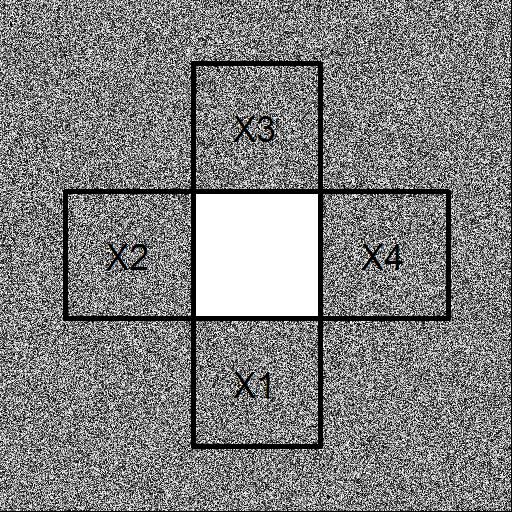}
 \caption{\label{fig:prediction} Block of missing values}
 \end{figure}

Now suppose that image $Z$ has a rectangular block of missing values. Without loss of generality, assume that the rectangular block of missing values is of size $(K-1)\times(K-1)$. Furthermore, in each border, $X^{(l)}$ ,$l=1,2,3,4$, is defined as a block of information of $Z$ of size $K \times K$, such as appears in Figure \ref{fig:prediction}.
 Also assume that for all $l$, $X^{(l)}$ is represented by a AR-2D model of the form
 \begin{equation*}
 X^{(l)}_{r,s}=\phi^{(l)} _{1}X^{(l)}_{r-1,s}+\phi^{(l)}_{2}X^{(l)}_{r,s-1}+\phi^{(l)}_{3}X^{(l)}_{r-1,s-1}+\varepsilon^{(l)}_{r,s},\quad l=1,2,3,4.
 \end{equation*}
 where $\phi^{(l)}_{1}$, $\phi^{(l)}_{2}$, and $\phi^{(l)}_{3}$ are estimated using the block $X^{(l)}$ for $l=1,2,3,4$, respectively. Then the prediction model is
 \begin{equation*}
 \widehat{X}^{(l)}_{r+i,s+j}=\left\{\begin{array}{lcc}
 \widehat{\phi}^{(l)}_{1}\widehat{X}^{(l)}_{r+i-1,s+j}+\widehat{\phi}^{(l)}_{2}\widehat{X}^{(l)}_{r+i,s+j-1}+\widehat{\phi}^{(l)}_{3}\widehat{X}^{(l)}_{r+i-1,s+j-1} &;& (r+i,s+j)\not\in A^{(l)}\\
 X^{(l)}_{r+i,s+j} &;& (r+i,s+j)\in A^{(l)}\end{array}\right.,
 \end{equation*}
 where $A^{(l)}$ is the index set for which $X^{(l)}$ is known and $i,j=1,\dots,K$. The prediction algorithm is the following

 \begin{algorithm}[h!!]
 \caption{Prediction Algorithm.}\label{alg:pred}
 {\bf Input}: An image $Z$ with a missing block, and $K$.
 
 {\bf Output}: Image $Z$ without missing values.
 \begin{algorithmic}[1]
 \State Get a sub-image $X$ of $Z$ of size $3K\times3K$, so that the missing data is in the center of $X$.
 \State Get $X^{(l)}$, for $l=1,2,3,4$, and reverse the order of the rows in $X^{(3)}$ and the columns in $X^{(4)}$, i.e. $X^{(3)}_{i,j}=X^{(3)}_{K+1-i,j}$, and $X^{(4)}_{i,j}=X^{(4)}_{i,K+1-j}$
 \State Compute $\widehat{\phi}^{(l)}_{1}$, $\widehat{\phi}^{(l)}_{2}$ and $\widehat{\phi}^{(l)}_{3}$ for $l=1,2,3,4$.
 \State Let $K_2=K$.
 \While{ $K_2>0$. }
 \For {$j=1$ {\bf until} $j=K-1$ }
 \State Compute: \begin{eqnarray*}
 X_{K+1,K+j} &=& \widehat{\phi}^{(1)}_{1}X_{K,K+j}+\widehat{\phi}^{(1)}_{2}X_{K+1,K+j-1}+\widehat{\phi}^{(1)}_{3}X_{K,K+j-1} \\
 X_{K+j,K+1} &=& \widehat{\phi}^{(2)}_{1}X_{K+j-1,K+1}+\widehat{\phi}^{(2)}_{2}X_{K+j,K}+\widehat{\phi}^{(2)}_{3}X_{K+j-1,K} \\
 X_{2K-1,K+j} &=& \widehat{\phi}^{(3)}_{1}X_{2K,K+j}+\widehat{\phi}^{(3)}_{2}X_{2K-1,K+j-1}+\widehat{\phi}^{(3)}_{3}X_{2K,K+j-1} \\
 X_{K+j,2K-1} &=& \widehat{\phi}^{(4)}_{1}X_{K+j-1,2K-1}+\widehat{\phi}^{(4)}_{2}X_{K+j,2K}+\widehat{\phi}^{(4)}_{3}X_{K+j-1,2K}
 \end{eqnarray*}
 \Comment For those points that the estimation is repeated consider the average of both estimations. These points are obtained for $j=1$ and $j=K-1$.
 
 {\bf end for}
 \EndFor
 \State Put $K_2=K_2-2$ and $K=K+1$.
 
 \EndWhile
 {\bf end while}
 \State Replace the NA values of $Z$ by $X$.
 
 \State {\bf Return} $Z$.
 \end{algorithmic}
 \end{algorithm}

%-------------------------------------------------------------------------------------------------
% REFERENCES
%-------------------------------------------------------------------------------------------------

\end{document}